\documentclass[a4paper,12pt]{article}

\usepackage{latexsym}
\usepackage{epsfig}
\usepackage{cite}
\usepackage{bm}
\input{colordvi.tex}

\usepackage{amsmath}
\usepackage{amsfonts}
\usepackage{amssymb}
\usepackage{graphicx}

\newcommand{\bi}{\begin{itemize}}
\newcommand{\ei}{\end{itemize}}
\newcommand{\beq}{\begin{equation}}
\newcommand{\eeq}{\end{equation}}
\newcommand{\bea}{\begin{eqnarray}}
\newcommand{\eea}{\end{eqnarray}}

\def\r{\rangle}

\def\N{{\cal N}}

\def\f{f}

\def\zetarad{\zeta_{\rm r}}
\def\i{{\rm inf}}

\def\k{{\bm k}}

\def\tOm{{\tilde\Omega}}
\def\curv{\sigma}
\def\f{f_c}
\def\r{r}

\def\z{z}
\def\s{s}
\def\ef{\varepsilon_f}

\def\tf{\tilde f}
\def\tg{\tilde g}
\def\SG{{\hat S}}

\begin{document}

\begin{titlepage}

\begin{center}

{\Large \bf  
Primordial Trispectrum from Isocurvature Fluctuations 
}

\vskip .45in

{\large
David Langlois$^1$ and
Tomo Takahashi$^2$
}

\vskip .45in

{\em
$^1$ APC (CNRS-Universit\'e Paris 7), 
10, rue Alice Domon et L\'eonie Duquet, 75205 Paris Cedex 13, France \\
\vspace{0.2cm}
$^2$Department of Physics, Saga University, Saga 840-8502, Japan 
  }

\end{center}

\vskip .4in

\begin{abstract}

  We study non-Gaussianity generated by adiabatic and isocurvature
  primordial perturbations.  We first obtain, in a very general
  setting, the non-linear perturbations, up to third order, for an
  arbitrary number of cosmological fluids, going through one or
  several decay transitions.  We then apply this formalism to the
  mixed curvaton and inflaton model, allowing for several decay
  channels. We compute the various contributions to the bispectrum and
  trispectrum resulting from adiabatic and isocurvature perturbations,
  which are correlated in general.  By investigating some hybrid decay
  scenario, we show that significant non-Gaussianity of adiabatic and
  isocurvature types can be generated without conflicting with the
  present isocurvature constraints from the power spectrum. In
  particular, we find cases where non-Gaussianity of isocurvature
  origin can dominate its adiabatic counterpart, both in the
  bispectrum and in the trispectrum.

\end{abstract}

\end{titlepage}

\clearpage

\setcounter{page}{1}

\section{Introduction}

Observations of cosmic density fluctuations, such as cosmic microwave
background (CMB) and large scale structure, are our only accessible
windows onto phenomena that occured in the early Universe. It is
therefore crucial to extract as much information as possible from
these observations, as a way to test or constrain various early
universe scenarios (see e.g. \cite{Langlois:2010xc} for recent lecture
notes).
 
This is the reason why primordial non-Gaussianity has been the subject
of an intense study recently, as it would provide invaluable
information beyond the power spectrum.  In particular, a detection of
primordial non-Gaussianity would rule out the simplest models of
inflation (with a single field), which generate only very small
non-Gaussianity.

Observations of the CMB anisotropies by the WMAP
satellite~\cite{Komatsu:2010fb} have set the present limit $f_{\rm
  NL}^{\rm local} =32\pm 21$ (68\,\% CL) [or $-10 < f_{\rm NL}^{\rm
  local} < 74$ (95\,\% CL)] on the non-linearity parameter $f_{\rm
  NL}^{\rm local} $, which characterizes the amplitude of the
bispectrum, for the local shape\footnote{ Three types of $f_{\rm NL}$
  have been discussed in the literatures.  Other two types and their
  constraints are $ -214 < f_{\rm NL}^{\rm equil} < 266 $ for the
  equilateral type and $ - 410< f_{\rm NL}^{\rm orthog} < 6 $ for the
  orthogonal type (95\,\% C.L.) \cite{Komatsu:2010fb}.
}.  Although Gaussian fluctuations are still consistent with current
observations, this constraint suggests that primordial fluctuations
may deviate from Gaussianity since the central value of this range is
somewhat away from zero.  The Planck satellite should be able to
confirm or infirm such level of non-Gaussianity. If Planck detects
significant local non-Gaussianity, then early Universe scenarios with
additional fields, such as a curvaton \cite{curvaton} or a modulaton
\cite{Dvali:2003em,Kofman:2003nx,Vernizzi:2003vs,Matsuda:2009yt,Langlois:2009jp,Kawasaki:2009hp},
would become natural alternatives to the simplest inflation scenarios.

As soon as one considers models with multiple scalar fields (several
inflatons or spectator fields beside the inflaton), one must take into
account the possibility that some (baryon or dark matter) isocurvature
perturbations may be generated, in addition to the usual adiabatic
fluctuations. The amplitude of such isocurvature modes, which could be
partially or fully correlated with the adiabatic one, is now severely
constrained by observations of the CMB power spectrum.  However,
complementary information, or constraints, on isocurvature modes can
be obtained from non-Gaussianity.

Non-Gaussianity in models with isocurvature fluctuations has been
studied in several works
\cite{Kawasaki:2008sn,Langlois:2008vk,Kawasaki:2008pa,Hikage:2008sk,Kawakami:2009iu,Langlois:2010dz}.
In particular, a general treatment for the primordial bispectrum which
allows various decay scenarios has been given in
\cite{Langlois:2010dz} recently.  Although the works mentioned above
have mainly studied the bispectrum (except \cite{Kawakami:2009iu}),
the trispectrum will also become, in the near future, a major target
since combined information from the bispectrum and the trispectrum is
a powerful way to discriminate between scenarios that generate local
non-Gaussianity, as discussed in detail in \cite{Suyama:2010uj}.  In
light of this, a general treatment of non-Gaussianity in models with
isocurvature fluctuations for the trispectrum is worth investigating
and is the main topic of this paper. More specifically, in the present
work, we extend the general formalism given in \cite{Langlois:2010dz}
to third order, then apply it to the mixed curvaton and inflaton
scenario~\cite{Langlois:2004nn,Moroi:2005kz,Moroi:2005np,Ichikawa:2008iq,Lemoine:2009is}.

The organization of this paper is as follows.  In the next section, we
present the general formalism to describe the curvature and
isocurvature perturbations, up to third order.  Then, in
Section~\ref{sec:curvaton}, we apply this formalism to the mixed
curvaton inflaton scenarios and compute the bispectra and trispectra
generated by the adiabatic and isocurvature modes in these scenarios.
Some quantitative discussions are also given. The final section is
devoted to the conclusion.  Throughout this paper, we set the reduced
Planck energy scale $M_{\rm pl}$ to be unity.

\section{General formalism}\label{sec:formalism}

In this section, we review and extend the general formalism introduced
in \cite{Langlois:2010dz} to describe the adiabatic and isocurvature
perturbations generated in scenarios with a cosmological transition
due to the decay of some species of particles, which we call $\curv$.
Explicit results up to second order were given in
\cite{Langlois:2010dz}, whereas we extend here the analysis up third
order, in order to be able to compute the relevant trispectra.

The purpose of the present analysis is to determine, at the non-linear
level, the perturbations of an arbitrary number of fluids, after a
cosmological transition where one of the fluid, labelled $\curv$,
decays. For any fluid $A$, one can define, in a covariant way
\cite{Langlois:2005ii,Langlois:2005qp} (see also the recent review
\cite{Langlois:2010vx}), a fully non-linear curvature perturbation
covector as \beq
\label{zeta_a}
\zeta_\mu^A=\nabla_\mu\N+\frac{\nabla_\mu\rho_A}{3(\rho_A+P_A)} \;,
\eeq where $\rho_A$ and $P_A$ are energy density and pressure for the
fluid A and $\N\equiv \int d\tau \nabla_\mu u^\mu/3$ is the number of
e-folds along the fluid worldlines (with proper time $\tau$) with
$u^\mu$ being the four-velocity of the fluid.

If $w_A\equiv P_A/\rho_A$ is constant, which will be assumed in the
following as we will later consider only relativistic species
($w=1/3$) and nonrelativistic species ($w=0$), the covector
(\ref{zeta_a}) can be written as the total gradient of \beq
 \label{defzeta}
  \zeta_{_A} = \delta \N +
\frac{1}{3(1+w_{_A})} \ln\left( \frac{\rho_{_A}}{\bar
    \rho_{_A}} \right)\; ,
 \eeq
 where $\delta \N\equiv \N-\bar{\N}$ and the barred quantities are
 defined in a reference spacetime, which is strictly homogeneous and
 isotropic. The fully non-linear perturbation $\zeta_A$, which
 coincides with the definition of \cite{Lyth:2004gb}, is conserved on
 large scales for a non-interacting fluid.

We  also introduce  the non-linear isocurvature perturbation of any fluid $A$, with respect to the radiation fluid, as
\beq
\label{iso_pert}
S_A\equiv 3(\zeta_A-\zeta_r),
\eeq
where $\zeta_r$ is the curvature perturbation of the radiation fluid.

Let us now focus on  the transition due to the decay.
In the sudden decay approximation, which we adopt here, the decay  takes place on the  hypersurface characterized by the condition
\beq
H_{\rm d}=\Gamma_\curv\, ,
\eeq
where $H_{\rm d}$ is the Hubble parameter at the decay and $\Gamma_\curv$ is 
the decay rate of $\curv$.
Since $H$ depends only on  the {\it total} energy density, the decay hypersurface is a hypersurface of uniform total energy density, with $\delta \N_{\rm d}=\zeta$, where $\zeta$ is the global curvature perturbation. 
The equality between the sum of all energy densities,  before the decay and  after the decay, thus reads
\beq
\label{bilan_decay}
\sum_A\bar{\rho}_{A-}e^{3(1+w_A)(\zeta_{A-}-\zeta)}=\bar{\rho}_{\rm decay}=\sum_B\bar{\rho}_{B+}e^{3(1+w_B)(\zeta_{B+}-\zeta)},
\eeq
where the subscripts $-$ and $+$ denote quantities defined, respectively, {\it before} and  {\it after} the transition. 
In the above formula, we have used the non-linear energy densities of the individual fluids, which can be expressed in terms of their curvature perturbation $\zeta_A$ by inverting the expression (\ref{defzeta}).

\subsection{Before the decay}
The first equality  in (\ref{bilan_decay}) implies
\beq
\sum_A \Omega_{A}e^{\beta_A(\zeta_{A-}-\zeta)}=1, \qquad \beta_A\equiv 3(1+w_A), 
\eeq
where we have introduced the relative abundances
$\Omega_A\equiv {\bar\rho_{A-}}/\bar\rho_{\rm decay}$, defined just {\it before} the decay. 

The above relation  determines $\zeta$ as a function of the $\zeta_{A-}$. 
Expanding it up to third order, one finds
\beq
\label{zeta_3}
\zeta=\sum_A\lambda_A\, \left[\zeta_{A-}+\frac{\beta_A}{2}\left(\zeta_{A-}-\zeta\right)^2
+\frac{\beta_A^2}{6}\left(\zeta_{A-}-\zeta\right)^3
\right] \, ,
\eeq
with  the coefficients
\beq
\label{lambda_A}
\lambda_A\equiv\frac{\tOm_A}{\tOm}, \qquad \tOm_A\equiv (1+w_A)\,  \Omega_A, \quad \tOm\equiv\sum_A\tOm_A\, .
\eeq
The global perturbation $\zeta$ appears on both sides of  the relation (\ref{zeta_3}), but one can use it iteratively to determine, order by order, the expression of $\zeta$ in terms of all the $\zeta_{A-}$, up to third order.

\subsection{After the decay}
In general, the  fluid $\curv$ can decay into various species, and it is convenient to introduce  
the relative branching ratio $\gamma_{A\curv}$, defined  as
\begin{equation}
\gamma_{A\curv}  \equiv \frac{\Gamma_{A\curv}}{\Gamma_{\curv}},\qquad \Gamma_\curv \equiv  \sum_A \Gamma_{A\curv},
\end{equation}
where $\Gamma_{A\curv}$ is the decay rate to $A$. Therefore, the energy density for any species $A$, just  after the decay of $\curv$,  
is simply given by
\begin{equation}
\rho_{A+} = \rho_{A-} + \gamma_{A\curv} \rho_\sigma.
\end{equation}
This relation, which is fully non-linear, can be reexpressed, upon using (\ref{defzeta}), in the form 
\beq
\label{bilan_A}
e^{\beta_A(\zeta_{A+}-\zeta)}=(1-f_A)e^{\beta_A(\zeta_{A-}-\zeta)}+f_Ae^{\beta_\sigma(\zeta_{\sigma -}-\zeta)} \, ,
\eeq
where the parameter
\beq
f_A\equiv \frac{\gamma_{A\sigma}\Omega_\sigma}{\Omega_A+\gamma_{A\sigma}\Omega_\sigma}\, 
\eeq
represents the fraction of the fluid $A$ that has been created by the decay. 

Expanding the above expression (\ref{bilan_A}) up to third order, and using (\ref{zeta_3}), one gets
\begin{eqnarray}
\label{zetaA+}
\zeta_{A+}&=&\sum_B T_{A}^{\ B}\left[\zeta_{B-}+\frac{\beta_B}{2}\left(\zeta_{B-}-\zeta\right)^2
+\frac{\beta_B^2}{6}\left(\zeta_{B-}-\zeta\right)^3
\right] 
\cr
&& -\frac{\beta_A}{2}\left(\zeta_{A+}-\zeta\right)^2
-\frac{\beta_A^2}{6}\left(\zeta_{A+}-\zeta\right)^3,
\end{eqnarray}
with the coefficients
\begin{eqnarray}
T_{A}^{\ A}&=& f_A\left(1-\frac{\beta_\curv}{\beta_A}\right)\lambda_A+(1-f_A),
\label{T1}
\\
T_{A}^{\ \curv}&=& f_A\left(1-\frac{\beta_\curv}{\beta_A}\right)\lambda_\curv+f_A\frac{\beta_\curv}{\beta_A},
\label{T2}
\\
T_{A}^{\ C}&=& f_A\left(1-\frac{\beta_\curv}{\beta_A}\right)\lambda_C\, , \quad C\neq A,\curv\, .
\label{T3}
\end{eqnarray}

At second order, after substituting the first order expression for $\zeta$ that follows from  (\ref{zeta_3}), Equation~(\ref{zetaA+}) yields
\beq
\label{zeta_2ndorder}
\zeta_{A+}=\sum_BT_A^{\ B}\zeta_{B-}+\sum_{B, C}U_A^{BC}\zeta_{B-}\zeta_{C-},
\eeq
with 
\begin{eqnarray*}
U_A^{BC}&\equiv&\frac12 \left[\sum_E\beta_E T_A^E(\delta_{EB}-\lambda_B)(\delta_{EC}-\lambda_C)-\beta_A(T_{AB}-\lambda_B)(T_{AC}-\lambda_C)
\right].
\end{eqnarray*}
Note that the above expression for $U_A^{BC}$ corresponds to the symmetrized version (with respect to the two indices $B$ and $C$) of the expression given in 
 \cite{Langlois:2010dz}. This does not change the expansion  (\ref{zeta_2ndorder}) since $U$ is 
 contracted with a symmetric term.

At third order, (\ref{zetaA+}) yields, after using  (\ref{zeta_3}) up to second order, 
\beq
\label{zeta_3rdorder}
\zeta_{A+}=\sum_BT_A^{\ B}\zeta_{B-}+\sum_{B, C}U_A^{BC}\zeta_{B-}\zeta_{C-}+
\sum_{B, C, D}V_A^{BCD}\zeta_{B-}\zeta_{C-} \zeta_{D-},
\eeq
with 
\begin{eqnarray*}
V_A^{BCD}&\equiv&-\frac12 \sum_{E,F}\beta_E T_{AE}(\delta_{EB}-\lambda_B)\lambda_F\beta_F (\delta_{FC}-\lambda_C) (\delta_{FD}-\lambda_D)
\cr
&& +\frac16 \sum_E\beta_E^2 T_{AE}(\delta_{EB}-\lambda_B)(\delta_{EC}-\lambda_C)(\delta_{ED}-\lambda_D)
\cr
&&
-\beta_A(T_{AB}-\lambda_B)\left[U_A^{CD}-\frac12\sum_E\beta_E\lambda_E (\delta_{EC}-\lambda_C)(\delta_{ED}-\lambda_D)\right]
\cr
&&
-\frac16\beta_A^2(T_{AB}-\lambda_B)(T_{AC}-\lambda_C)(T_{AD}-\lambda_D)\, .
\end{eqnarray*}
The above expression is the main result of this section. It provides a systematic computation of the post-decay curvature perturbations for all fluids in a very general setting. For scenarios with several decay transitions, the perturbations can be obtained by combining the various expressions of the type (\ref{zeta_3rdorder}) for each transition. This was done explicitly, up to second order, for a scenario with two curvatons in  \cite{Langlois:2010dz}. In the present work, we will consider only  single curvaton scenarios and thus use (\ref{zeta_3rdorder}) only once.

\section{Mixed curvaton  and inflaton scenario} \label{sec:curvaton}

\def\xr{r}
\def\xc{x_c}
\def\fc{f_c}
We now apply the general formalism  given in the previous section to the mixed curvaton scenario, where fluctuations of both the curvaton and the inflaton are taken into account\footnote{
The adiabatic (curvature) perturbation in such a scenario 
has been investigated in 
\cite{Langlois:2004nn,Moroi:2005kz,Moroi:2005np,Ichikawa:2008iq,Lemoine:2009is}.
}
. 
 
Although CDM isocurvature fluctuations and their non-Gaussianity have
been discussed in several works
\cite{Moroi:2002rd,Lyth:2003ip,Beltran:2008ei,Langlois:2008vk,Moroi:2008nn,Takahashi:2009cx},
it is usually assumed that all of the CDM (or baryons for baryon
isocurvature perturbations) is produced either before the curvaton
decay or as a product of the curvaton decay.  However, one can also
envisage (as in \cite{Kawasaki:2008pa,Langlois:2010dz}) intermediate
situations, where the CDM is produced both before and during the
curvaton decay.  Such cases can be treated in the formalism discussed
in the previous section and lead, as we will see, to interesting
observational consequences.

Here we consider a simplified scenario with three initial species:
radiation ($r$), CDM ($c$) and a curvaton ($\curv$).  We also assume
that the potential for the curvaton is quadratic\footnote{
The self-interacting curvaton model which include some non-quadratic potential 
has been studied in 
\cite{Enqvist:2005pg,Huang:2008zj,Enqvist:2009zf,Enqvist:2009ww,Byrnes:2010xd}.
}, and we thus treat 
$\curv$ as a pressureless fluid, which then decays  into radiation and CDM.

\subsection{Perturbations after the decay}
\subsubsection{Transfer matrix}
In the present case, we find that the linear transfer matrix, whose  coefficients are defined in \eqref{T1}--\eqref{T3}, is given  by  
\beq
\label{T}
T= 
	\left( \begin{array}{ccccc}
	1-\xr && \xc && \xr-\xc  \\
	0 && 1-\fc && \fc  \\
	0  && 0 && 0 
	\end{array} \right), 
	\quad \xr\equiv \frac{f_r}{\tOm}, \quad x_c\equiv \frac14\Omega_c\, \xr,
\eeq
where the order  of the species is $(r,c,\curv)$. We have used the definitions (\ref{lambda_A}) for the coefficients $\lambda_A$, which leads in particular to $\lambda_r=4(1-\tOm^{-1})$ (since $\tOm=1+\Omega_r/3$ in our scenario with two nonrelativistic species).

In the following, we will assume $\Omega_c\ll 1$, since the decay  must occur deep in the radiation dominated era\footnote{Note that, although $\Omega_c$ is assumed to be very small, it cannot be neglected in the  expression for $\f$  because $\gamma_{c\curv}$ or $\Omega_\sigma$ can be very small, and $\f$ can take any value between $0$ and $1$.}. 
As a consequence, we assume  $\lambda_c=0$ and, therefore, $x_c=0$. This also implies $\tOm=(4-\Omega_\sigma)/3$.

In order to compare easily our results with the existing literature, it will  be convenient to rewrite  the parameter $\xr$ as
\beq
\r\equiv \xi \, {\tilde r}\, ,
\eeq
where the quantity 
\beq
\xi \equiv 
\frac{f_r}{\Omega_\curv}=
 \frac{\gamma_{r \, \curv}}{1-(1- \gamma_{r \, \curv}) \Omega_\curv}
\eeq
can be interpreted as  the  efficiency of the energy transfer from the curvaton into radiation ($\xi=1$ if the curvaton decays only into radiation, i.e. $\gamma_{r\sigma}=1$), 
and 
\beq
{\tilde r}=  \frac{3\Omega_\curv}{4-\Omega_\curv}
\eeq
 is the familiar coefficient that appears in the literature on the curvaton, which coincides with our $r$ only if  $\xi=1$.

\subsubsection{Nonlinear perturbations}
It is now straightforward to obtain the post-decay  curvature perturbations for each of the remaining fluid, up to third order, by using our   general result 
(\ref{zeta_3rdorder}). 

The expression for  the CDM curvature perturbation reads
\beq
\zeta_{c+}=\zeta_{c}+\frac13 f_c (S_{\curv}-S_c)+\frac16 f_c (1-f_c)\left(S_\curv-S_c\right)^2+\frac{1}{18} \f(1-3\f+2\f^2)(S_\curv-S_c)^3,
\label{iso_mixed3}
\eeq
where, on the right-hand side, we have substituted $\zeta_\curv-\zeta_c=(S_\curv-S_c)/3$ and  the pre-decay subscript  is implicit for all quantities.

The above expression, which turns out to be  valid also for an arbitrary value of $x_c$, can in fact be derived much more rapidly by using directly (\ref{bilan_A}) for the CDM fluid. Indeed, since $\beta_c=\beta_\curv=3$, one gets
\beq
\zeta_{c+}=\zeta+\frac13\ln\left[(1-\f)e^{3(\zeta_c-\zeta)}+\f e^{3(\zeta_\curv-\zeta)}\right]=\zeta_c +\frac13\ln\left[1+\f \left(e^{S_\curv-S_c}-1\right)\right],
\eeq
and expanding the last expression gives immediately (\ref{iso_mixed3}).

The curvature perturbation for radiation is much more complicated in the general case. In the limit $x_c=0$, the full expression reduces to
\begin{eqnarray}
\zeta_{r+}&=& \zeta_{r-}+\frac{\r}{3} S_{\curv-}+\frac{\r}{18}\left[3
-4r +\frac{2r}{\xi} - \frac{r^2}{\xi^2}  \right] S_{\curv-}^2 \cr
   &+&\frac{\r}{162}\left[ 9+18 (1-2 \xi ) \frac{r}{\xi}+4  \left(8 \xi ^2-6 \xi -3\right) \frac{r^2}{\xi ^2} +2  (6 \xi -1)\frac{r^3}{\xi^3}+3 \frac{r^4}{\xi^4} \right]S_{\curv-}^3\qquad 
\label{curv_mixed3}
\end{eqnarray}
In the limit  $\xi=1$, one recovers the expression given in \cite{Sasaki:2006kq}.

\subsection{Primordial adiabatic and isocurvature perturbations}

In order to determine the statistical properties of the primordial
adiabatic and isocurvature perturbations, defined at the onset of the
standard cosmological era, i.e. after the curvaton decay, one needs to
relate the perturbation of the curvaton {\it fluid} to the
fluctuations of the curvaton {\it field}, generated during inflation.

For a massive curvaton without self-interaction, the relation between the isocurvature perturbation $S_\curv$ and the fluctuation $\delta\curv$ is simply given by
the non-linear expression
 \beq
 e^{S_\curv}
 =
 \left( 1+\frac{\delta\curv}{\bar\curv} \right)^2 \,. \label{rhorhobarcurv} 
 \end{equation}
 This result can be obtained by writing the (non-linear) energy density of  the oscillating curvaton defined on the spatially flat
hypersurfaces, characterized by $\delta N=\zeta_r$ when the curvaton is still  subdominant:
\beq
\rho_\curv = m^2 \curv^2=m^2 \left( \bar\curv+\delta\curv \right)^2= \bar\rho_\curv e^{3(\zeta_\curv-\zeta_r)}=\bar\rho_\curv e^{S_\curv}\, .
\eeq
 
Expanding the expression (\ref{rhorhobarcurv}) up to  third
order,  and using the conservation of $\delta\curv/\curv$ in a quadratic potential, we obtain
 \beq
 \label{S_G}
S_\curv= \SG-\frac14 \SG^2+\frac{1}{12}\SG^3\, ,
\eeq
where the quantity 
 \beq
 \SG \equiv 2\frac{\delta\curv_*}{\bar\curv_*}\, 
 \eeq
is Gaussian.

For simplicity, we now restrict our analysis to the situation where 
\beq
\zeta_{c-}=\zeta_{r-}\equiv\zeta_\i\,,
\eeq
by assuming that the CDM and radiation perturbations, {\it before} the curvaton decay, depend only on the inflaton fluctuations. This means that, effectively,  
 there are only two independent degrees of freedom from the inflationary epoch, $\zeta_\i$ and $\SG$.

Substituting  (\ref{S_G}) into (\ref{curv_mixed3}) and (\ref{iso_mixed3}) then yields the primordial curvature perturbation 
\beq
\label{zetarad}
\zetarad =\zeta_\i+\z_1 \SG+\frac12 \z_2 \SG^2+\frac16 \z_3 \SG^3\, ,
\eeq
with 
\begin{eqnarray}
 z_1&=&  \frac{\r}{3}\,, \qquad  
 z_2=\frac{\r}{18}\left(3 -8\r +\frac{4r}{\xi} -2\frac{r^2}{\xi^2} \right), 
\\
z_3&=&\frac{\r^2}{54}\left(\frac{6 r^3}{\xi ^4}+\frac{24 r^2}{\xi ^2}-\frac{4 r^2}{\xi ^3}-\frac{48 r}{\xi }-\frac{15
   r}{\xi ^2}+64 r+\frac{18}{\xi }-36\right) \,,
\end{eqnarray}
and the primordial isocurvature perturbation
\beq
\label{S_c}
S_c=\s_1 \SG+\frac12 \s_2 \SG^2+\frac16\s_3 \SG^3\, ,
\eeq
with 
\begin{eqnarray}
 s_1&=&\f-r, \qquad s_2=\frac{1}{6}\left(3 \f (1-2\f)+\frac{2 r^3}{\xi ^2}-\frac{4 r^2}{\xi }+8 r^2-3 r\right)\, ,
\\
s_3&=&-\frac12\f^2(3-4\f)-\frac{\r^2}{18}\left(\frac{6 r^3}{\xi ^4}+\frac{24 r^2}{\xi ^2}-\frac{4 r^2}{\xi ^3}-\frac{48 r}{\xi }-\frac{15
   r}{\xi ^2}+64 r+\frac{18}{\xi }-36\right)\,.\qquad
\end{eqnarray}
These expressions will be useful to determine the power spectrum, the bispectrum and the trispectrum,
which can be probed by observations.

\subsection{Power spectrum}

The power spectrum for the total curvature perturbation is given, 
according to (\ref{zetarad}), by
 \beq
 \label{finalzetar}
  {\cal P}_{\zetarad}={\cal P}_{\zeta_\i}+\frac{r^2}{9}{\cal P}_{\SG} \equiv (1+\lambda){\cal P}_{\zeta_\i}\equiv \Xi^{-1}\frac{r^2}{9}{\cal P}_{\SG},
 \eeq
where $\lambda$ is the ratio between the curvaton and inflaton contributions,
\begin{equation}
\lambda \equiv \frac{(r^2/9) {\cal P}_{\SG}}{ {\cal P}_{\zeta_\i}},
\end{equation}
and $\Xi=1/(1+\lambda^{-1})$ represents the fraction of the power spectrum due to the curvaton contribution. 
Since the contribution from the inflaton (for standard slow-roll single field inflation)  can be written as 
\begin{equation}
{\cal P}_{\zeta_\i} = \frac{1}{2\epsilon} {\cal P}_{\delta \phi},
\end{equation}
with ${\cal P}_{\delta \phi}$ being the power spectrum of the inflaton 
fluctuations $\delta \phi$ and $\epsilon = (1/2) (d V/d\phi)^2 / V^2$ being the slow-roll parameter.
The parameter $\lambda$ is explicitly given by 
\begin{equation}
\label{lambda}
\lambda = \frac{8 r^2 \epsilon }{9\sigma_\ast^2}.
\end{equation}
Thus $\lambda$ depends on the curvaton model parameters, as well as  on the inflation model. 

We can also write down the power spectrum for the isocurvature fluctuations. 
Using Eq.~\eqref{S_c}, we find
\begin{equation}
{\cal P}_{S_c} = (f_c - r)^2 {\cal P}_{\SG}.
\end{equation}
In our model, both curvature and isocurvature perturbations depend on the curvaton fluctuations. Therefore, the two types of perturbations are   correlated, 
 as quantified by  the correlation 
coefficient:
\begin{equation}
\label{corr_ad_is}
{\cal C}
= \frac{ {\cal P}_{S_c, \zeta_r} }{\sqrt{ {\cal P}_{S_c} {\cal P}_{\zeta_r}}} =\ef \sqrt{\Xi}, \qquad 
\ef\equiv {\rm sgn} (f_c -r) .
\end{equation}
In the pure curvaton limit ($\Xi\simeq 1$), adiabatic and isocurvature perturbations are either fully correlated, if $\ef>0$, or fully anti-correlated, if $\ef<0$.
In the opposite limit ($\Xi\ll 1$), the correlation vanishes. 
For intermediate values of  $\Xi$, the correlation is only partial, as can also be  obtained  in multifield inflation
\cite{Langlois:1999dw}.

Finally, it is convenient to introduce the ratio of the isocurvature power spectrum with respect to the adiabatic one
\begin{equation}
\label{eq:alpha}
\alpha = \frac{{\cal P}_{S_c}}{{\cal P}_{\zeta_r}} = 9\left(1-\frac{f_c}{ r}\right)^2 \,\Xi \, .
\end{equation}
This ratio can be strongly constrained by cosmological observations, but the precise limits depend on the assumed level of correlation between the isocurvature and adiabatic perturbations (since the impact of isocurvature perturbations on the observable power spectrum depends crucially on this correlation, as illustrated in ~\cite{Langlois:2000ar}). 
Constraints on $\alpha$ from the latest data have been published only for the uncorrelated and fully correlated cases. 
In terms of  the parameter $a\equiv \alpha/(1+\alpha)$, the limits (based on WMAP+BAO+SN data) given in  \cite{Komatsu:2010fb} 
are\footnote{Our notations differ from those of \cite{Komatsu:2010fb}: our $a$ corresponds to their $\alpha$ and our fully {\sl correlated} limit corresponds to their fully {\sl anti-correlated} limit, because their definition of the correlation has the opposite sign 
(see also \cite{Komatsu:2008hk} for a more detailed discussion)
.}
\beq
a_0<0.064\quad (95 \% {\rm CL}), \qquad a_{1}< 0.0037 \quad (95 \% {\rm CL})\,,
\eeq
respectively for the uncorrelated case ($\Xi=0$) and for the fully correlated case ($\Xi=1$).

According to the expression (\ref{eq:alpha}), the observational constraint $\alpha \ll 1$ can be satisfied if $|\f-r|\ll r$ (which includes the case   $\f=1$ with  $r\simeq  1$)
or if  $\Xi \ll 1$, i.e. the curvaton contribution to the observed power spectrum is very small.


\subsection{Bispectrum}

We now discuss  the three-point functions for our curvature and isocurvature perturbations, summarizing the results obtained in \cite{Langlois:2010dz}.

Let us start by considering, in  the general case of an arbitrary number of observable quantities $X^I$, all possible bispectra, defined by 
\beq
\langle X^{I}_{\k_1} X^{J}_{\k_2} X^{K}_{\k_3} \rangle 
= (2 \pi)^3 \delta (\Sigma_i \k_i) B^{IJK}(k_1, k_2, k_3)\,.
 \eeq
 We then assume  that the $X^I$ can be written, up to second order, in the form
\beq
\label{X_I_2}
X^I= N^I_a \phi^a+\frac12 N^{I}_{ab} \delta\phi^a \delta\phi^b + \dots
\eeq
(with implicit summation over the indices $a$ and $b$), 
where the $N_a^I$ and $N_{ab}^I$ are arbitrary background-dependent coefficients and 
 the $\delta\phi^a$ are Gaussian fluctuations, generated during inflation and characterized by their 
 two-point correlation functions 
\beq
\label{P_ab}
 \langle \delta\phi^a (\k) \delta\phi^b (\k')\rangle=
 (2\pi)^3 \, P^{ab}(k) \, \delta(\k + \k') \,.
 \eeq
 Substituting the decomposition (\ref{X_I_2}) into the left hand side, and using (\ref{P_ab}), 
 one finds 
 \cite{Lyth:2005fi}
 \begin{eqnarray}
 B^{IJK}(k_1, k_2, k_3)&=&N_a^I N_b^J N_{cd}^K P^{ac}(k_1) P^{bd}(k_2)+N_a^IN_{bc}^JN_d^KP^{ab}(k_1) P^{cd}(k_3)
 \cr
 && 
 +N^I_{ab}N^J_cN^K_d P^{ac}(k_2)P^{bd}(k_3).
 \end{eqnarray}
 
 In our particular case, we have only two observables $X^I=\{\zeta_r, S_c\}$, corresponding to  two indices which we denote $I=\{\zeta, S\}$. Moreover, since  there is only one Gaussian degree of freedom, $\SG$, in the non-linear expansions for $\zeta_r$ and $S_c$, respectively (\ref{zetarad}) and (\ref{S_c}), the bispectra reduce to 
\begin{eqnarray}
 B^{IJK}(k_1, k_2, k_3)=
 b_{NL}^{I, JK}  P_\SG(k_2) P_\SG(k_3)+b_{NL}^{J, KI}  P_\SG(k_1) P_\SG(k_3)+b_{NL}^{K, IJ}   P_\SG(k_1)P_\SG(k_2),\quad 
 \end{eqnarray}
 with 
 \beq
b_{NL}^{I, JK} \equiv N^I_{(2)} N^J_{(1)} N^K_{(1)},  
\eeq
where $N^\zeta_{(2)}=z_2$, $N^S_{(2)}=s_2$, $N^\zeta_{(1)}=z_1$, $N^S_{(1)}=s_1$, respectively.

In order to compare these coefficients with the usual parameter\footnote{
All the nonlinear coefficients that we introduce are of the local type and we drop the superscript ``local" for simplicity.
} $f_{NL}$  defined in the purely adiabatic case, one must remember that $f_{NL}$ is proportional to the bispectrum of $\zeta$ divided by the square of the power spectrum. The analogs of $f_{NL}$ can therefore be defined  by dividing  the coefficient $b_{NL}^{I, JK}$ by the square  of the ratio $P_\zeta/P_{\SG}=z_1^2/\Xi$:
\beq
\tf_{NL}^{I,JK}\equiv \frac65 f_{NL}^{I, JK} \equiv \frac{\Xi^2}{z_1^4} \, b_{NL}^{I, JK} \, .
\eeq
Taking into account the fact that the last two indices can be permuted, this leads to   six different coefficients,  
explicitly given by the expressions
\begin{eqnarray}
\tf_{NL}^{\zeta, \zeta\zeta}&=&\frac{z_2}{z_1^2}\, \Xi^2,  \quad \tf_{NL}^{\zeta, \zeta S}=\frac{s_1 z_2}{z_1^3}\, \Xi^2,  \quad  \tf_{NL}^{S, \zeta\zeta}=
\frac{s_2}{z_1^2}\, \Xi^2,
\\
\tf_{NL}^{\zeta, S S}&=&\frac{s_1^2 z_2}{z_1^4}\, \Xi^2,  \quad \tf_{NL}^{S, \zeta S}= \frac{s_1 s_2}{z_1^3}\, \Xi^2,  \quad \tf_{NL}^{S, SS}= \frac{s_1^2 s_2}{z_1^4}\, \Xi^2\, . 
\end{eqnarray}

It is worth noting that all the coefficients are related via  the two rules
\beq
\label{rules}
 f_{NL}^{I, JS}= \frac{s_1}{z_1} f_{NL}^{I, J\zeta}, 
 \qquad f_{NL}^{S, IJ}= \frac{s_2}{z_2} f_{NL}^{\zeta, IJ}.
\eeq
Therefore, the hierarchy between the parameters can be deduced from the value of the  ratios $s_1/z_1$ and 
$s_2/z_2$, which are given in the small $r$ limit (assuming $\xi=1$) by the simple expressions
\beq
\label{ratios_sz}
\frac{s_1}{z_1}=\sqrt{\frac{\alpha}{\Xi}}\, =3\left(\frac{\f}{\r}-1\right)\,,\quad
\frac{s_2}{z_2}\simeq \frac{3\f(1-2\f)}{\r}-3, \qquad (r\ll 1, \xi=1)\,.
\eeq

Observational constraints on isocurvature non-Gaussianities are given in  \cite{Hikage:2008sk}, for an isocurvature perturbation  of the form $S=\SG+f_{NL}^{\rm (iso)}\SG^2$, 
where $\SG$ is Gaussian. The relations between  the non-linear  parameter $f_{NL}^{\rm (iso)}$ and 
the parameters defined above are the following:  $\tf_{NL}^{S, SS}= 2 f_{NL}^{\rm (iso)}\alpha^2$, $\tf_{NL}^{S, \zeta S}=2f_{NL}^{\rm (iso)} \alpha^{3/2} |{\cal C}|$ and 
$\tf_{NL}^{S, \zeta\zeta} =2 f_{NL}^{\rm (iso)}\alpha\,  {\cal C}^2$, where $\alpha$ and ${\cal C}$ are respectively defined in \eqref{eq:alpha} and \eqref{corr_ad_is}.

\subsection{Trispectrum}
We now turn to the novel part of this work, which consists of the study of the trispectrum.  

Let us first consider the general  case of  several observables $X^I$ and let us introduce  the trispectra  $T^{IJKL}$ defined from the  connected
four-point correlation functions as
\beq
\langle X^{I}_{\k_1} X^J_{\k_2} X^{K}_{\k_3} X^L_{\k_4}\rangle_c 
\equiv 
(2 \pi)^3 \delta (\Sigma_i \k_i) T^{IJKL}(\k_1, \k_2, \k_3, \k_4)\,.
 \eeq
If the observables can  be written, up to third order,  in the form
\beq
\label{X_I}
X^I= N^I_a \delta\phi^a+\frac12 N^{I}_{ab} \delta\phi^a \delta\phi^b + \frac16 N^{I}_{abc} \delta\phi^a \delta\phi^b\delta\phi^c+\dots
\eeq
one finds that the trispectra are given by \cite{Byrnes:2006vq}
\begin{eqnarray}
T^{IJKL}(\k_1, \k_2, \k_3, \k_4)&=& 
N^I_{a_1a_2 a_3} N^J_{b}N^K_cN^L_d P^{a_1b}(k_2)P^{a_2c}(k_3) P^{a_3d}(k_4) + 3\ {\rm perms}
\cr
\cr
&+& N^I_{a_1a_2} N^J_{b_1b_2}N^K_cN^L_d\left[P^{a_1c}(k_3)P^{b_1d}(k_4) P^{a_2b_2}(k_{13})
\right.
\cr 
&& \left.
+P^{b_1c}(k_3)P^{a_1d}(k_4) P^{a_2b_2}(k_{14})\right] + 5 \ {\rm perms} \, ,
\end{eqnarray}
with the notation $k_{13}\equiv |\k_1+\k_3|$, etc.

In our particular case, where there is only one Gaussian degree of freedom at  the non-linear level, the trispectra reduce to 
\begin{eqnarray}
T^{IJKL}(\k_1, \k_2, \k_3, \k_4)&=& 
t_{NL}^{I,JKL} P_\SG(k_2)P_\SG(k_3) P_\SG(k_4) + 3\ {\rm perms}
\cr
\cr
&+& \hat t_{NL}^{IJ,KL}  \left[P_\SG(k_3)P_\SG(k_4) P_\SG(k_{13})
+P_\SG(k_3)P_\SG(k_4) P_\SG(k_{14})\right] + 5 \ {\rm perms}\, ,  \qquad 
\end{eqnarray}
with 
\beq
t_{NL}^{I,JKL}\equiv N^I_{(3)} N^J_{(1)}N^K_{(1)} N^L_{(1)}, \qquad \hat t_{NL}^{IJ,KL} =N^I_{(2)} N^J_{(2)}N^K_{(1)}N^L_{(1)},
\eeq
where $N^\zeta_{(3)}=z_3$ and  $N^S_{(3)}=s_3$, in analogy with the notations introduced previously.
Taking into account  the symmetries under permutations of the indices, one finds, for two observables ($I=\{\zeta, S\}$), 8 different parameters $t_{NL}^{I,JKL}$ and 9 parameters $\hat t_{NL}^{IJ,KL}$.

In order to facilitate the comparison with the parameters $\tau_{NL}$ and $g_{NL}$, which have been defined in the purely adiabatic case by dividing the trispectrum by the cube of the power spectrum, it is convenient to rescale the above parameters and
 to introduce the coefficients
\beq
\tau_{NL}^{IJ,KL} \equiv\frac{N^I_{(2)} N^J_{(2)}N^K_{(1)}N^L_{(1)}}{z_1^6}\Xi^3\, ,
\eeq
which depend only on the second order non-linearities, and the coefficients
\beq
\tg_{NL}^{I,JKL}\equiv \frac{54}{25}g_{NL}^{I,JKL}\equiv \frac{N^I_{(3)} N^J_{(1)}N^K_{(1)} N^L_{(1)}}{z_1^6}\Xi^3 , 
\eeq
which depend on the third order non-linearities.

For the purely adiabatic parameters, we obtain
\begin{eqnarray}
\tau_{\rm NL}^{\zeta\zeta,\zeta\zeta} 
& = &   \frac{\z_2^2}{\z_1^4}\, \Xi^3\,,
\\
 \tg_{\rm NL}^{\zeta,\zeta\zeta\zeta} 
 & = &  \frac{\z_3}{\z_1^3}\, \Xi^3\,,
\end{eqnarray}
which exactly coincide with the usual $\tau_{\rm NL}$ and $(54/25) g_{\rm NL}$. 
The present constraints on these parameters, assuming the data do not contain any isocurvature contribution, are~\cite{Smidt:2010sv}
\begin{eqnarray*}
-7.4 <10^{-5} g_{NL} < 8.2 \  (95\%\,  {\rm CL}) , \qquad  -0.6  < 10^{-4} \tau_{NL} < 3.3 \ (95\%\,  {\rm CL})\,.
\end{eqnarray*}
Similarly for the (purely) isocurvature mode, we have
\begin{eqnarray}
\tau_{\rm NL}^{SS, SS}
& = &   \frac{\s_1^2\s_2^2}{\z_1^6}\, \Xi^3\,,
\\
 \tg_{\rm NL}^{S, SSS} 
 & = &  \frac{\s_1^3\s_3}{\z_1^6}\, \Xi^3\,.
\end{eqnarray}

In addition to the purely adiabatic or isocurvature 
nonlinear coefficients, 
we also find the following {\it cross-correlated}  
nonlinear coefficients:
\begin{eqnarray}
\tau_{\rm NL}^{\zeta\zeta, \zeta S}&=&\frac{\s_1\z_2^2}{\z_1^5}\, \Xi^3, \quad \tau_{\rm NL}^{\zeta S, \zeta \zeta} = \frac{\z_2\s_2}{\z_1^4}\, \Xi^3\,,
\\
 \tg_{\rm NL}^{S, \zeta\zeta\zeta} &=& \frac{\s_3}{\z_1^3}\, \Xi^3, \qquad  \tg_{\rm NL}^{\zeta, \zeta\zeta S} 
 =  \frac{\s_1\z_3}{\z_1^4}\, \Xi^3\,,
  \end{eqnarray}
\begin{eqnarray}
\tau_{\rm NL}^{\zeta\zeta,  S S}
& = &  \frac{\s_1^2\z_2^2}{\z_1^6}\, \Xi^3, \quad
\tau_{\rm NL}^{\zeta S, \zeta S}= \frac{\s_1\z_2\s_2}{\z_1^5}\, \Xi^3, \quad 
\tau_{\rm NL}^{SS, \zeta\zeta}= \frac{\s_2^2}{\z_1^4}\, \Xi^3\,,
    \\
  \tg_{\rm NL}^{\zeta, \zeta  S S}
  &=& \frac{s_1^2z_3}{z_1^5} \, \Xi^3, \quad  
  \tg_{\rm NL}^{S,S \zeta\zeta} = 
  \frac{s_1 s_3}{z_1^4} \, \Xi^3\,,
  \\
\tau_{\rm NL}^{\zeta S, S S}
& = &   \frac{\s_1^2\z_2\s_2}{\z_1^6}\, \Xi^3, \qquad 
\tau_{\rm NL}^{SS, \zeta S}
 =    \frac{\s_1\s_2^2}{\z_1^5}\, \Xi^3,
   \\
 \tg_{\rm NL}^{\zeta, S S S} 
 & = &  \frac{\s_1^3\z_3}{\z_1^6}\, \Xi^3, \quad 
  \tg_{\rm NL}^{S, \zeta S S } 
  =   \frac{\s_1^2\s_3}{\z_1^5}\, \Xi^3\,.
  \end{eqnarray}

These nonlinear coefficients can be related by very simple rules, in analogy with the rules obtained for the bispectrum coefficients. The first rule is that the replacement of an index $\zeta$ by $S$ on the {\it right hand side} of the comma corresponds to a rescaling by $s_1/z_1$:
\beq
g_{NL}^{I,JKS}=\frac{s_1}{z_1} g_{NL}^{I,JK\zeta}, \qquad \tau_{NL}^{IJ,KS}= \frac{s_1}{z_1}\tau_{NL}^{IJ,K\zeta} \,. 
\eeq
The replacement of $\zeta$ by $S$ on the {\it left hand side} of the comma leads to a rescaling by $s_2/z_2$ for the $\tau_{NL}^{IJ,KL}$ and a rescaling by $s_3/z_3$ for the $g_{NL}^{I,JKL}$:
\beq
g_{NL}^{S,JKL}=\frac{s_3}{z_3}g_{NL}^{\zeta, JKL}, \qquad 
\tau_{NL}^{IS,KL}=\frac{s_2}{z_2}\tau_{NL}^{I\zeta,KL}.
\eeq
These rules will be extremely useful in the subsequent discussion, enabling us to establish hierarchies between 
the 17 nonlinear  coefficients defined above. 

Moreover, there are  consistency relations between the trispectrum parameters $\tau_{NL}^{IJ,KL}$ and the bispectrum parameters $f_{NL}^{I, JK}$. For example, one finds
\beq
\tau_{NL}^{IJ,\zeta\zeta}=\Xi^{-1}f_{NL}^{I, \zeta\zeta}
f_{NL}^{J, \zeta\zeta}\,, 
\eeq
relating the trispectrum parameters to the bispectrum ones, which generalizes the purely adiabatic consistency relation $\tau_{NL}=(6f_{NL}/5)^2/\Xi$, discussed in  \cite{Suyama:2010uj}. 

\subsection{Quantitative discussion}

In the pure curvaton scenario,  when CDM is created before 
the curvaton decay (i.e. $\f=0)$,
 large isocurvature fluctuations (correlated with the adiabatic ones)
are generated and they turn out to be too large to be compatible with current data.  One way out 
is to  consider the mixed inflaton and curvaton scenario,  where the inflaton fluctuations also contribute to the observed power spectrum, in addition to the curvaton ones. The  isocurvature fluctuations are then ``diluted," (by the $\Xi$ factor) and  can thus be made consistent with observations,  as   studied in \cite{Moroi:2002rd,Langlois:2008vk,Moroi:2008nn,Takahashi:2009cx}.

At the same time, since  
non-Gaussianity in these models originates from the curvaton sector, one  can naively expect that non-Gaussianity will also be small if the curvaton contribution to the power spectrum  is small.
Thus it seems that large non-Gaussianity is difficult to  realize  
without conflicting with the isocurvature constraint, although 
 $f_{\rm NL} > {\cal O}(10)$ is still possible while satisfying isocurvature constraint \cite{Moroi:2008nn,Takahashi:2009cx}.
However, in the models investigated in the present work,  we also include the possibility for  CDM to  be created 
both from the curvaton decay and from some pre-decay epoch. As we will see  explicitly below, this  leads to 
very interesting consequences in the parameter space.

\subsubsection{Bispectrum parameters}
\begin{figure}[ht]
\begin{center}
\resizebox{150mm}{!}{\includegraphics{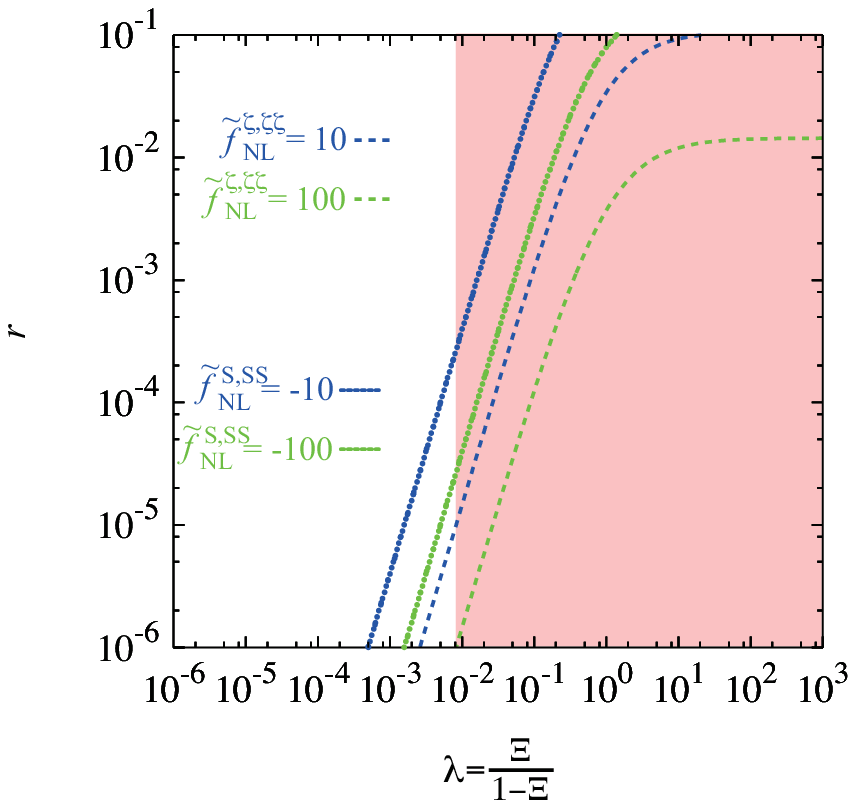}
\includegraphics{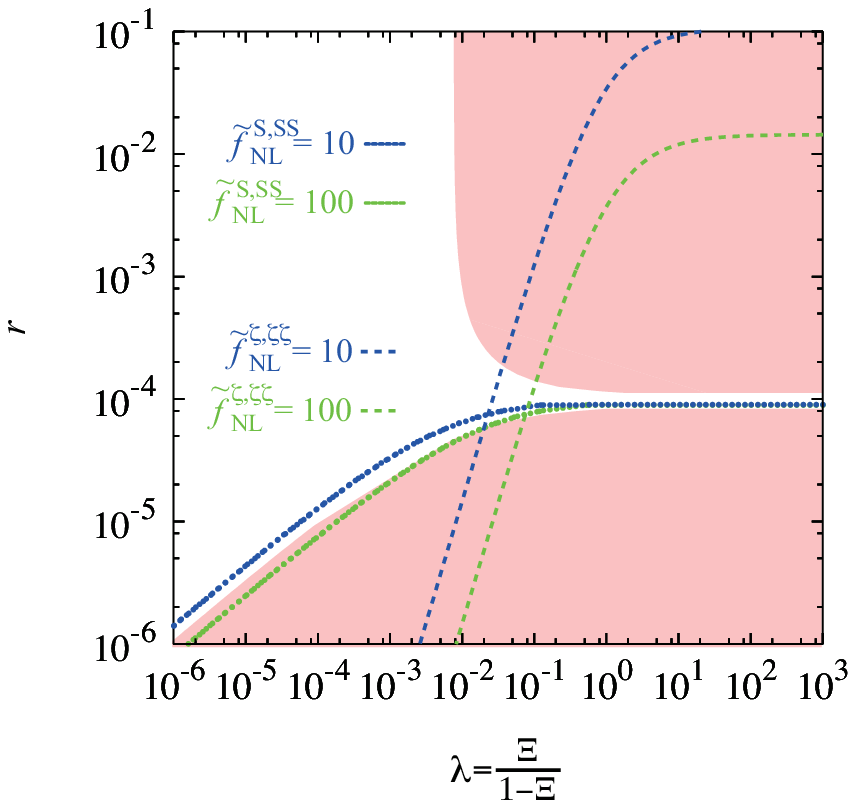}}
\caption{
Contours for the bispectrum coefficients $\tilde{f}_{\rm NL}^{\zeta,\zeta\zeta}$ and $\tilde{f}_{\rm NL}^{S,SS}$
in the $\lambda$--$r$ plane in the cases with $f_c = 0$ (left) and $10^{-4}$ (right).
The regions $a > 0.064$ are shaded. 
}
\label{fig:fnl}
\end{center}
\end{figure}

 In Fig.~\ref{fig:fnl}, we have plotted the contours of $\tf_{\rm NL}^{\zeta,\zeta\zeta}$ and 
$\tf_{\rm NL}^{S, SS}$ in the $\lambda$--$r$ parameter plane, for two values of $\f$ and assuming $\xi=1$.   

We recall that the parameter $\lambda$, defined in (\ref{lambda})  as the ratio between the curvaton and inflaton contributions to the power spectrum, is directly 
related to the parameter $\Xi$, which we have prefered to use in the analytical expressions for the nonlinear coefficients,  by the relation 
\beq
\lambda= \frac{\Xi}{1-\Xi}.
\eeq
In the limit $\lambda\ll 1$, one gets  $\Xi\simeq \lambda$, whereas $\Xi\simeq 1-\lambda^{-1}$ in  the limit $\lambda\gg 1$.

The constraint ($a_0 < 0.064$ at 95 \% C.L. ) on 
uncorrelated isocurvature mode from WMAP7
 is also used to  identify  regions still allowed  by the current data\footnote{ 
Although this limit, strictly speaking, applies  only for the region 
$\lambda \ll 1$, we have used the same limit for the intermediate region $\lambda\sim 1$ as well as for the region $\lambda\gg1$. In the latter region, we know that the real limit is more stringent when $\ef=1$ since it is given by the constraint on  $a_{1}$. But  constraints from the current data on $a_{-1}$ or $a_\Xi$ with intermediate values for $\Xi$ are not given in the literature, so we have prefered to use the limit on $a_0$ everywhere as indicative constraints on the parameter space. }.
For small values of $r$, the purely adiabatic parameter is given by 
\beq
\tf_{\rm NL}^{\zeta,\zeta\zeta}\simeq \frac{3}{2r}\, \Xi^2, 
\eeq
and the other $\tf_{NL}^{I,JK}$ can be deduced from it  by appropriate factors of  $s_1/z_1$ or $s_2/z_2$, according to (\ref{rules}). 

\begin{figure}[ht]
\begin{center}
\resizebox{80mm}{!}{\includegraphics{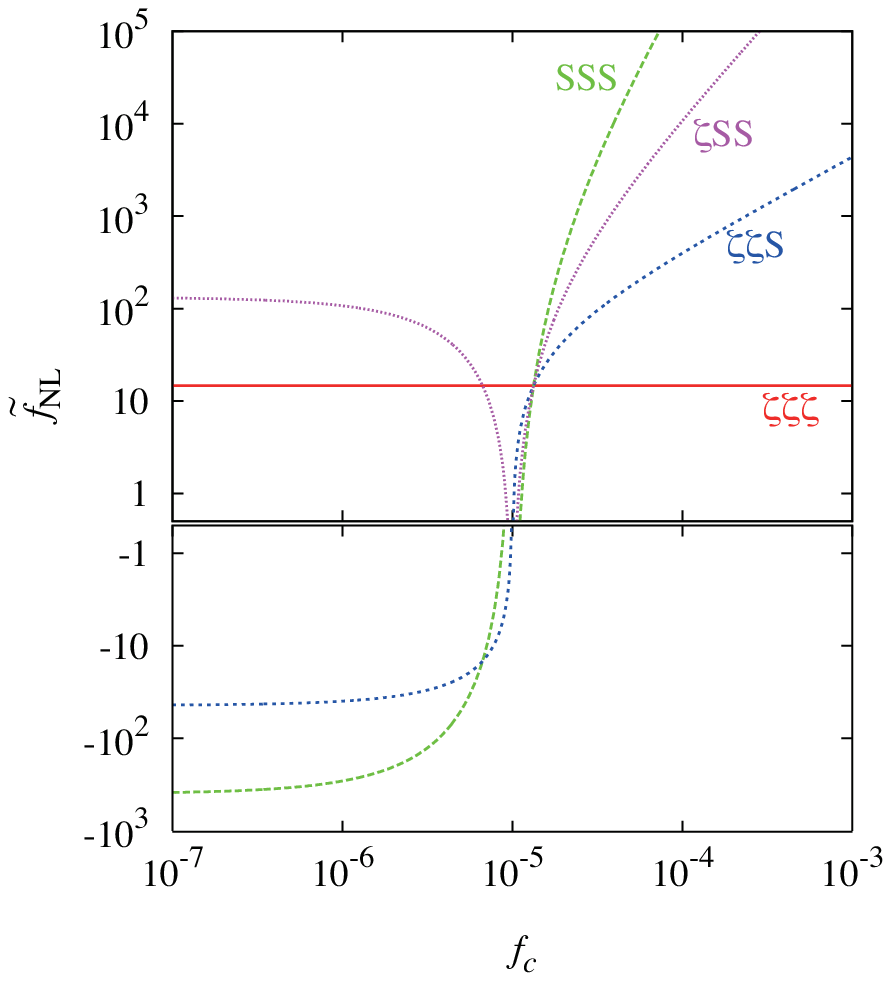}}
\caption{
Plots of the coefficients $\tf_{\rm NL}^{I,JK}$ as functions of $f_c$.
The indices $I,J,K$ are specified in the figure for each curve.
The other parameters are  $\xi=1$, $\lambda = 10^{-3} \simeq \Xi$ and $r=10^{-5}$.
}
\label{fig:fnl_fc}
\end{center}
\end{figure}

In Fig.~\ref{fig:fnl_fc}, we have plotted all the parameters  $\tf_{\rm NL}^{I,JK}$ as a function of $f_c$, for fixed (and small) values of $\Xi$ and $\r$. The figure shows how the parameters evolve from  the region $\f \ll r$ to the region $\f\gg \r$.
The figure clearly illustrates the various hierarchies between the non-linearity coefficients, which we discuss below.

In the case of $\f=0$, illustrated  in the first plot of Fig.~\ref{fig:fnl}, both  factors $s_1/z_1$ and
 $s_2/z_2$ reduce to $(-3)$. As a consequence, the $f_{NL}^{I,JK}$ are slightly enhanced with respect to $f_{\rm NL}^{\zeta,\zeta\zeta}$ by a factor $(-3)^{I_S}$ where $I_S$ is the number of indices $S$. In particular the pure isocurvature parameter is $(-27)$ times $f_{NL}$ as confirmed by the figures. 
 The constraint  on the isocurvature  power spectrum 
 imposes $\Xi\ll 1$ and the non-Gaussianity can therefore  be significant only if $r$ is sufficiently small to compensate the $\Xi^2$ suppression.

When $\f$ does not vanish, as 
illustrated in the second plot of Fig.~\ref{fig:fnl}, a new region appears in the parameter space, where the isocurvature constraint is satisfied even if $\Xi\simeq 1$. 
This requires a fine-tuning between $\f$ and $r$, more precisely, that of 
\beq
\f-r\simeq \ef \frac{\sqrt{\alpha}}{3} r \qquad (\Xi\simeq 1), 
\eeq
according to  \eqref{eq:alpha}.
  With respect to the purely adiabatic non-Gaussianity, the other types of non-Gaussianities are suppressed in general (except 
$f_{NL}^{S,\zeta\zeta}$, which could be of the same order of magnitude in the particular case $\xi\ll 1$ and ${\tilde r}\equiv r/\xi\sim 1$).

In the case $r \ll f_c  \ll 1$, another interesting feature appears in  the region  $\Xi \ll 1$. Indeed,   the isocurvature bispectrum amplitude $f_{\rm NL}^{S,SS}$ becomes larger than $f_{\rm NL}^{\zeta,\zeta\zeta}$, as pointed out in \cite{Langlois:2010dz}. The reason is that the factors $s_1/z_1$ and $s_2/z_2$ become $3\f/r$ in this region 
so that the $f_{NL}$ with adiabatic indices are suppressed with respect to the purely isocurvature non-Gaussianity. In this region, $f_{\rm NL}^{\zeta,\zeta\zeta}$ and $f_{\rm NL}^{S,SS}$ are 
respectively given by
\begin{eqnarray}
\tf_{\rm NL}^{\zeta,\zeta\zeta} 
\simeq
\frac{3}{2r} \Xi^2
\simeq 
\frac{r^3}{54f_c^4} \alpha^2, \qquad
 \tf_{\rm NL}^{S,SS} 
 \simeq 
 \frac{81 f_c^3}{2r^4} \Xi^2
 \simeq 
 \frac{\alpha^2}{2 f_c}, \qquad (r \ll f_c  \ll 1),
\end{eqnarray}
from which we can see that $\tf_{\rm NL}^{S,SS}$ is enhanced by the factor $(3\f/r)^3$ compared 
to $\tf_{\rm NL}^{\zeta,\zeta\zeta}$ and can be significant if $f_c$ is well below $\alpha^2$. 
In such a scenario, future observations would thus  detect non-Gaussianity from the isocurvature perturbations rather than  from the adiabatic ones. 
For the correlated 
nonlinear coefficients  $f_{\rm NL}^{I,JK}$,  we obtain
\beq
\tilde{f}_{\rm NL}^{\zeta, \zeta S} \simeq  \tilde{f}_{\rm NL}^{S,  \zeta\zeta} 
 \simeq  \displaystyle\frac{9f_c}{2r^2} \Xi^2,  \qquad  \qquad 
\tilde{f}_{\rm NL}^{\zeta, S S} \simeq  \tilde{f}_{\rm NL}^{S,  \zeta S}
 \simeq \displaystyle\frac{27}{2} \displaystyle\frac{f_c^2}{r^3} \Xi^2. \\
\eeq
Their value depends only on the total number of adiabatic indices. The hierarchy between 
 the non-linearity parameters depends only on the ratio $\f/\r$. 

In the opposite  limit where $f_c$ is much smaller than $r$ ($f_c \ll r \ll 1$), all
the non-linearity 
coefficients
  become independent  of $\f$. Once again, the amplitude of the purely isocurvature coefficient is enhanced (now  by a  factor $27$) 
with respect to the purely adiabatic one, but the sign is changed:
\beq
\tf^{\zeta,\zeta\zeta}_{\rm NL}\simeq \frac{3}{2r}\Xi^2\simeq \frac{\alpha^2}{54 r}, \qquad \tilde{f}_{\rm NL}^{S,SS} \simeq  - \displaystyle\frac{81}{2r} \Xi^2\, .
\eeq
The correlated 
nonlinear coefficients
 have  intermediate values,
\begin{eqnarray}
\tilde{f}_{\rm NL}^{\zeta,\zeta S} \simeq  \tilde{f}_{\rm NL}^{S, \zeta\zeta} 
 \simeq -  \displaystyle\frac{9}{2r} \Xi^2,  
 \qquad  
\tilde{f}_{\rm NL}^{\zeta, S S} \simeq  \tilde{f}_{\rm NL}^{S, \zeta S} 
 \simeq  \displaystyle\frac{27}{2r} \Xi^2.
\end{eqnarray}
All these values differ simply by powers of $(-3)$, since $s_1/z_1\simeq s_2/z_2\simeq -3$ in this limit. An interesting consequence is that the 
nonlinear coefficients with an odd number of isocurvature indices are 
negative.

\subsubsection{Trispectrum parameters}

\begin{figure}[ht]
\begin{center}
\resizebox{150mm}{!}{\includegraphics{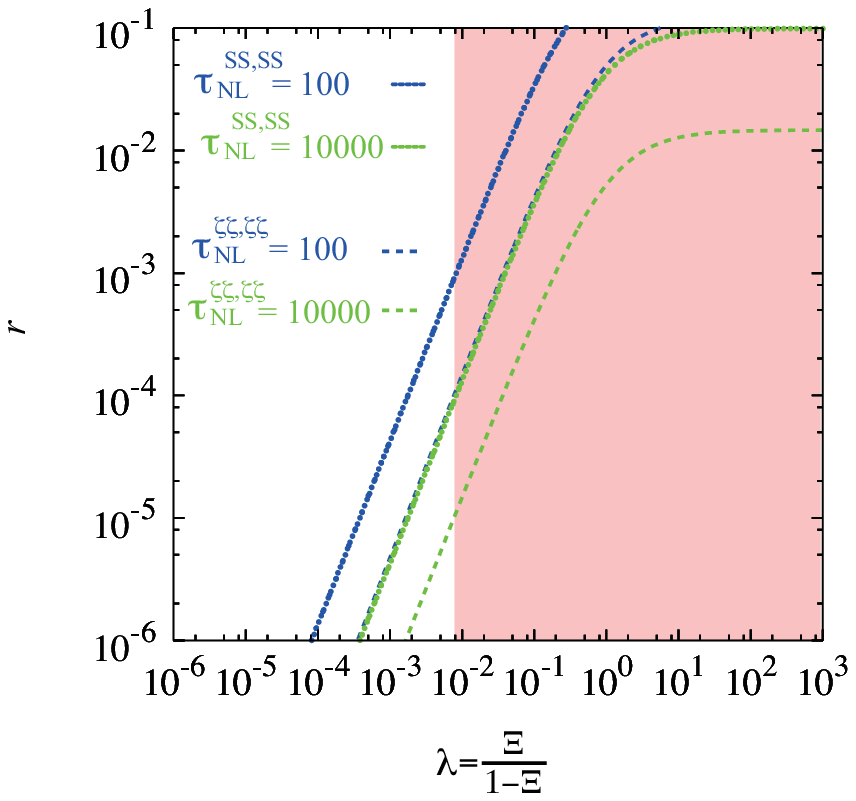}
\includegraphics{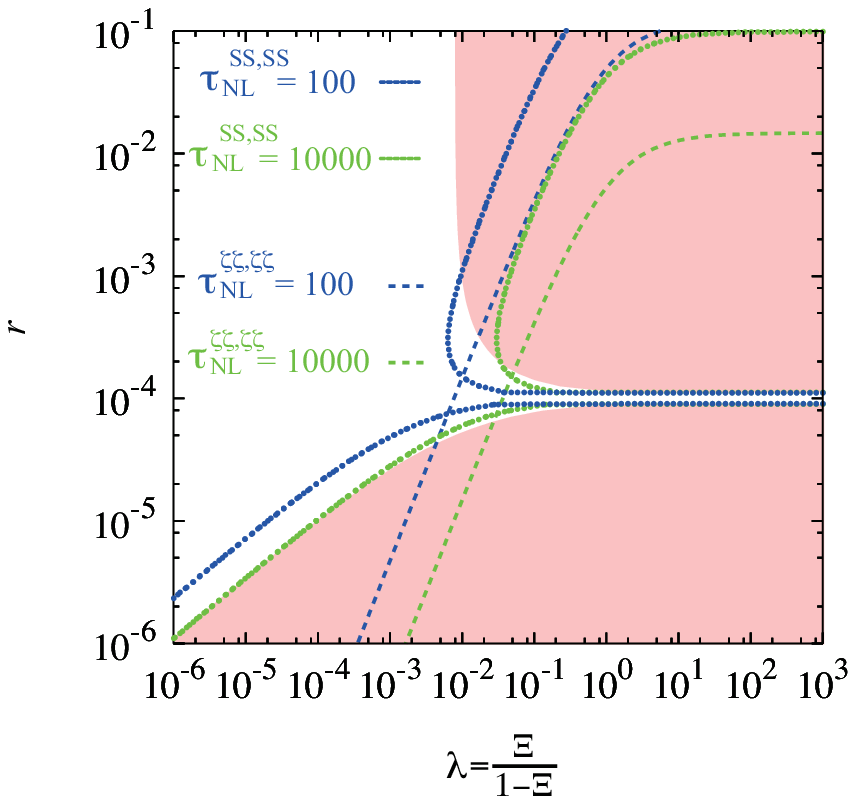}}
\caption{Contours for the trispectrum coefficients $\tau_{\rm NL}^{\zeta\zeta,\zeta\zeta}$ and $\tau_{\rm NL}^{SS,SS}$
in the $\lambda$--$r$ plane for the cases  $f_c = 0$ (left) and $10^{-4}$ (right). The regions $a> 0.064$ are shaded. 
}
\label{fig:taunl}
\end{center}
\end{figure}

\begin{figure}[ht]
\begin{center}
\resizebox{150mm}{!}{\includegraphics{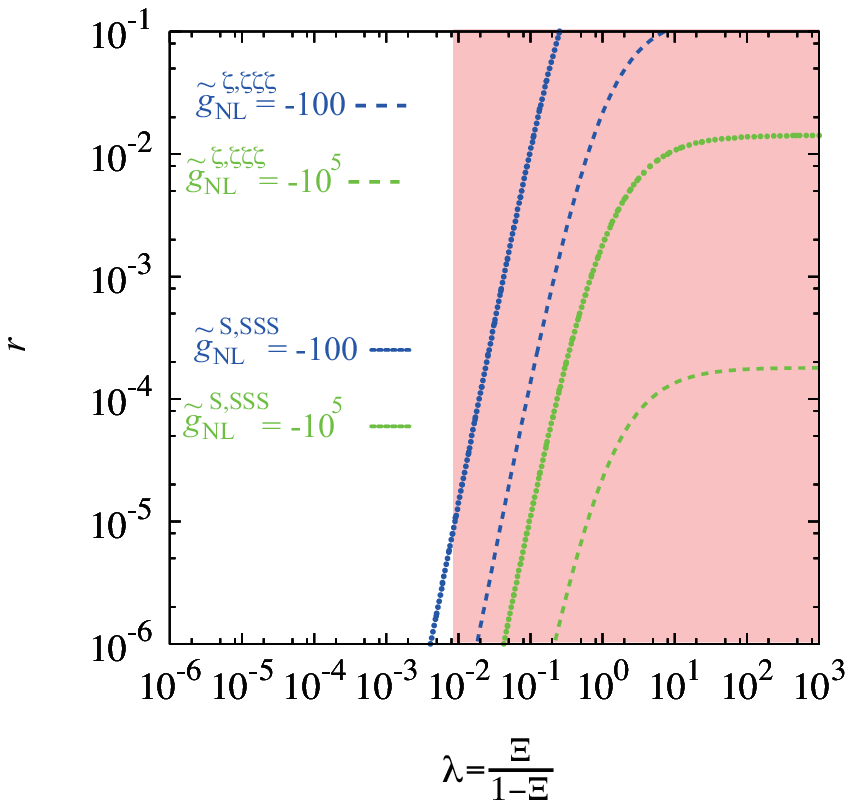}
\includegraphics{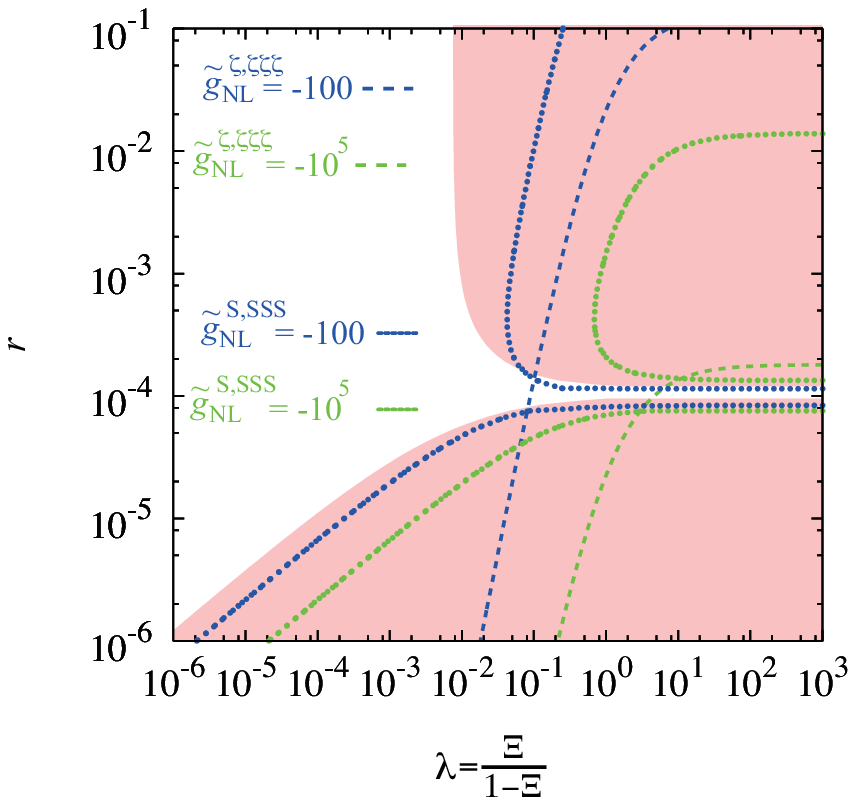}}
\caption{Contours for the trispectrum coefficients $\tilde{g}_{\rm NL}^{\zeta,\zeta\zeta\zeta}$ and 
$\tilde{g}_{\rm NL}^{S,SSS}$
in the $\lambda$--$r$ plane for the cases with $f_c = 0$ (left) and $10^{-4}$ (right).
The regions $a> 0.064$ are shaded. 
}
\label{fig:gnl}
\end{center}
\end{figure}

The contours of the purely adiabatic and isocurvature 
nonlinear coefficients
 $\tau_{\rm NL}^{\zeta\zeta,\zeta\zeta}$ and 
$\tau_{\rm NL}^{SS,SS}$ have been plotted in Fig.~\ref{fig:taunl}, while  the coefficients  
$\tilde{g}_{\rm NL}^{\zeta,\zeta\zeta\zeta}$ and 
$\tilde{g}_{\rm NL}^{S,SSS}$  are  depicted in Fig.~\ref{fig:gnl}. 
In these figures, the region excluded by current observations of the isocurvature  mode is again shaded. 

For small values of $\r$, the purely adiabatic trispectrum coefficients are (for $\xi=1$)
\beq
\tau_{\rm NL}^{\zeta\zeta,\zeta\zeta}
\simeq \frac{9}{4r^2}\Xi^3, \qquad 
\tg_{\rm NL}^{\zeta,\zeta\zeta\zeta}\simeq -\frac{9}{r}\Xi^3, \qquad (r\ll 1, \ \xi=1).
\eeq
They can be also related to the 
purely adiabatic nonlinear coefficient $\tf_{NL}\equiv \tf_{NL}^{\zeta,\zeta\zeta}$ by the following expressions \cite{Suyama:2010uj}:
\beq
\label{bi_tri_ad}
\tau_{\rm NL}^{\zeta\zeta,\zeta\zeta}=\tf_{NL}^2/\Xi, \qquad 
g_{\rm NL}^{\zeta,\zeta\zeta\zeta} \simeq -\frac{10}{3} f_{NL}\, \Xi\,.
\eeq
This implies that, when $\Xi\ll 1$, $\tau_{NL}^{\zeta\zeta,\zeta\zeta}$ is enhanced with respect to $\tf_{NL}^2$ whereas, by contrast, $g_{NL}^{\zeta,\zeta\zeta\zeta}$ is suppressed with respect to $f_{NL}$. 

For $\f=0$, all $\tau_{NL}^{IJ,KL}$ are slightly enhanced with respect to the adiabatic $\tau_{NL}$ by factors $(-3)^{I_S}$. In particular, $\tau_{NL}^{SS,SS}=81 \, \tau_{NL}$. The same conclusion applies to  the $\tg_{NL}^{I, JKL}$ since $s_3/z_3=-3$ for $\f=0$. 

In the fine-tuned region $\f\sim r$, the purely isocurvature 
nonlinear coefficients and most of the correlated coefficients  are suppressed with respect to the purely adiabatic ones (except in the case $\xi\ll 1$, as discussed earlier). However, since $s_3/z_3\simeq 3/2$, one finds 
\beq
\tg^{S,\zeta\zeta\zeta}_{NL}\simeq \frac32 \tg_{NL}^{\zeta,\zeta\zeta\zeta}, \qquad (\f\sim r, \ \Xi\simeq 1, \ \xi=1).
\eeq

In the region $r \ll f_c  \ll 1$ and $\Xi \ll 1$, where the bispectrum is dominated by its purely isocurvature component, we find the same conclusion for the trispectrum coefficients 
since $s_3/z_3\sim f^2/r^2$. The  largest coefficients are thus (for $\xi=1$)
\beq
\tau_{NL}^{SS,SS}\simeq \frac{729 \f^4}{4\r^6}\Xi^3\simeq \frac{\alpha^3}{4\f^2},\quad 
\tg_{NL}^{S,SSS}\simeq -\frac{2187 \f^5}{2\r^6}\Xi^3 \simeq-\frac{3\alpha^3}{2\f}  \qquad (r \ll f_c  \ll 1, \xi=1),
\eeq
and one finds the following relations with $f_{NL}^{S,SS}$:
\beq
\tau_{NL}^{SS,SS}\simeq  \left( \tf_{NL}^{S,SS}\right)^2 \alpha^{-1}, \qquad \tg_{NL}^{S,SSS}\simeq -3\alpha \tf_{NL}^{S,SS},
\eeq
which are very similar to their adiabatic counterparts (\ref{bi_tri_ad}). 

\begin{figure}[ht]
\begin{center}
\resizebox{80mm}{!}{\includegraphics{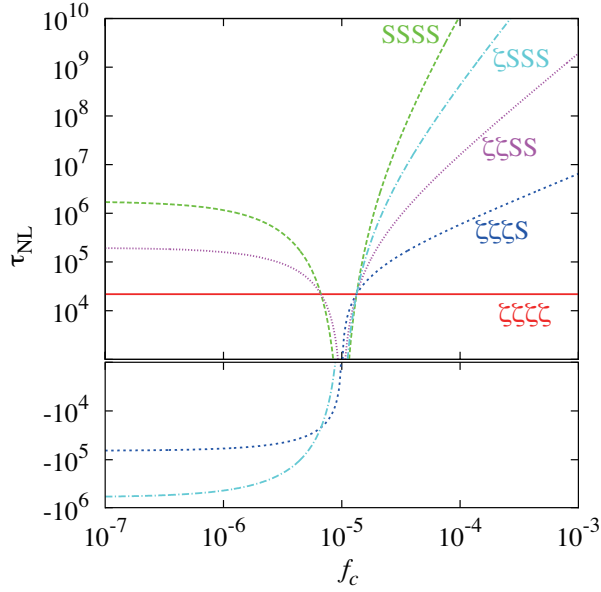}}
\caption{Plots of the coefficients $\tau_{\rm NL}^{IJ,KL}$ as functions of $f_c$.
The indices $I, J, K,L$ are specified in the figure for each curve. The
other parameters are  $\xi=1, \lambda = 10^{-3}$ and $r=10^{-5}$.}
\label{fig:taunl_fc}
\end{center}
\end{figure}

In Fig.~\ref{fig:taunl_fc}, we have plotted the evolution of the hierarchy between all the parameters $\tau_{\rm NL}^{IJ,KL}$ as the parameter $\f$ varies, $\r$ being kept fixed (with a small value). 
In the limit $r \ll f_c \ll 1$, one sees clearly that the purely isocurvature coefficient dominates, while  the other coefficients vary according to 
\beq
\tau_{\rm NL}^{IJ,KL}\simeq \left(\frac{\r}{3\f}\right)^{I_\zeta} \tau_{NL}^{SS,SS},
\qquad ( r \ll f_c \ll 1),
\eeq
where $I_\zeta$ is the number of adiabatic indices among the four indices of the coefficient.
In this opposite limit,  $f_c \ll r \ll 1$, we have
\beq
\tau_{\rm NL}^{IJ,KL}\simeq \left(-\frac{1}{3}\right)^{I_\zeta} \tau_{NL}^{SS,SS},
\qquad ( f_c \ll r \ll 1).
\eeq

\begin{figure}[ht]
\begin{center}
\resizebox{80mm}{!}{\includegraphics{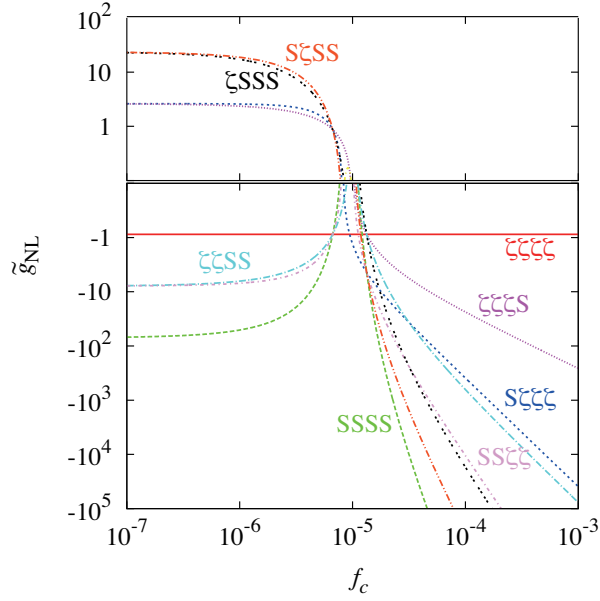}}
\caption{Plots of $\tilde{g}_{\rm NL}^{IJKL}$ as a function of $f_c$.
$I, J, K,L$ are specified in the figure for each line. Other parameters are fixed as $\xi=1, \lambda = 10^{-3}$ and $r=10^{-5}$.}
\label{fig:gnl_fc}
\end{center}
\end{figure}

In Fig.~\ref{fig:gnl_fc},  we have plotted the evolution of all the  $\tg_{\rm NL}^{I, JKL}$ as $\f$ varies. Once again, the two opposite limits have very distinctive hierarchies between the nonlinear coefficients. 
In the limit $r \ll f_c \ll 1$, $\tg_{NL}^{S,SSS} $ dominates. 
Moreover, 
\beq
\tg_{NL}^{\zeta,SSS} \simeq -\frac{243 f_c^3}{r^4}\,  \Xi^3\simeq \frac{2\r^2}{9\f^2} \, \tg_{NL}^{S,SSS}, 
\eeq
and all the other coefficients can be deduced by using the relations
\beq
  \tg_{\rm NL}^{S,JKL}\simeq \left(\frac{\r}{3\f}\right)^{I^{\hat 3}_\zeta} \tg_{NL}^{S,SSS}, \quad   \tg_{\rm NL}^{\zeta,JKL}\simeq \left(\frac{\r}{3\f}\right)^{I^{\hat 3}_\zeta} \tg_{NL}^{\zeta,SSS},\qquad ( r \ll f_c \ll 1),
\eeq
where $I^{\hat 3}_\zeta$ is the number of adiabatic indices among the three indices after the comma.
In the opposite limit,  $f_c \ll r \ll 1$, the hierarchy is simply
\beq
\tg_{\rm NL}^{I,JKL}\simeq \left(-\frac{1}{3}\right)^{I_\zeta} \tg_{NL}^{S,SSS},
\qquad ( f_c \ll r \ll 1),
\eeq
where $I_\zeta$ is the number of adiabatic indices among the four indices.

\section{Conclusion}

In this paper, we have investigated non-Gaussianity in models with isocurvature 
fluctuations, paying particular attention to the trispectrum. After presenting a general formalism 
to calculate bi- and tri-spectra of adiabatic, isocurvature and  correlated types,
allowing various decay scenarios, we have applied the formalism 
to the mixed curvaton and inflaton model.  

We have studied how the amplitude  of the non-linearity parameters, which consist of  
six  $f_{\rm NL}^{I,JK}$, nine $\tau_{\rm NL}^{IJ,KL}$ and  eight $g_{\rm NL}^{I,JKL}$,  depend on the parameters 
$r$, the curvaton contribution to the power spectrum $\Xi$ (or, equivalently, $\lambda$) and  $f_c$, which represents the fraction 
of dark matter produced from the curvaton decay. 
We have found  that large non-Gaussianity (in the bispectrum and trispectrum)  from both of adiabatic 
and isocurvature modes is possible in some cases without conflicting the isocurvature constraint from
the measured CMB power spectrum. We have also compared  the relative size of the non-linearity coefficients of various types as depicted in Figs.~\ref{fig:fnl_fc}, \ref{fig:taunl_fc}  and \ref{fig:gnl_fc}, and found that different regions 
in the parameter space correspond to very distinctive hierarchies between  the non-Gaussianity coefficients.

Observations of non-Gaussianity, and in particular of the trispectrum  have become an important 
goal in cosmology. From another perspective, isocurvature fluctuations are associated 
with the generation of dark matter and baryon asymmetry in the Universe.
Thus, non-Gaussianity from isocurvature fluctuations, if  detected in the future,  would give us a lot of insight into
the nature of dark matter, the mechanism of baryogenesis, and therefore into  high energy physics.

\section*{Acknowledgments}
We would like to thank Angela Lepidi for helpful comments on the manuscript. 
T.~T. would like to thank APC for the hospitality during the visit, where this work was finalized.
We also thank the organizers of ``Cosmology -- The Next Generation" (YKIS 2010), 
where preliminary discussions that led to the present work were initiated. 
This  work   is partially supported by the ANR (Agence Nationale de la
Recherche) grant ``STR-COSMO", ANR-09-BLAN-0157 (D.~L.) and  by the
Grant-in-Aid for Scientific research from the Ministry of Education,
Science, Sports, and Culture, Japan, No. 19740145 (T.~T.).


\begin{thebibliography}{100}


\bibitem{Langlois:2010xc}
  D.~Langlois,
  Lect.\ Notes Phys.\  {\bf 800}, 1 (2010)
  [arXiv:1001.5259 [astro-ph.CO]].


\bibitem{Komatsu:2010fb}
  E.~Komatsu {\it et al.},
  arXiv:1001.4538 [astro-ph.CO].
 
 
  \bibitem{curvaton}
  A.~D.~Linde and V.~F.~Mukhanov,
  Phys.\ Rev.\  D {\bf 56} (1997) 535
  [arXiv:astro-ph/9610219].
  K.~Enqvist and M.~S.~Sloth,
  Nucl.\ Phys.\  B {\bf 626}, 395 (2002)
  [arXiv:hep-ph/0109214];
  D.~H.~Lyth and D.~Wands,
  Phys.\ Lett.\  B {\bf 524}, 5 (2002)
  [arXiv:hep-ph/0110002];
  T.~Moroi and T.~Takahashi,
  Phys.\ Lett.\  B {\bf 522}, 215 (2001)
  [Erratum-ibid.\  B {\bf 539}, 303 (2002)]
  [arXiv:hep-ph/0110096].
 
  
\bibitem{Dvali:2003em}
  G.~Dvali, A.~Gruzinov and M.~Zaldarriaga,
  Phys.\ Rev.\  D {\bf 69}, 023505 (2004)
  [arXiv:astro-ph/0303591].
  
\bibitem{Kofman:2003nx}
  L.~Kofman,
  arXiv:astro-ph/0303614.
  

\bibitem{Vernizzi:2003vs}
  F.~Vernizzi,
  Phys.\ Rev.\  D {\bf 69}, 083526 (2004)
  [arXiv:astro-ph/0311167].

\bibitem{Matsuda:2009yt}
  T.~Matsuda,
  Class.\ Quant.\ Grav.\  {\bf 26}, 145011 (2009)
  [arXiv:0902.4283 [hep-ph]].

\bibitem{Langlois:2009jp}
  D.~Langlois and L.~Sorbo,
  JCAP {\bf 0908}, 014 (2009)
  [arXiv:0906.1813 [astro-ph.CO]].


\bibitem{Kawasaki:2009hp}
  M.~Kawasaki, T.~Takahashi and S.~Yokoyama,
  JCAP {\bf 0912}, 012 (2009)
  [arXiv:0910.3053 [hep-th]].

\bibitem{Kawasaki:2008sn}
  M.~Kawasaki, K.~Nakayama, T.~Sekiguchi, T.~Suyama and F.~Takahashi,
  JCAP {\bf 0811}, 019 (2008)
  [arXiv:0808.0009 [astro-ph]].



\bibitem{Langlois:2008vk}
  D.~Langlois, F.~Vernizzi and D.~Wands,
  JCAP {\bf 0812}, 004 (2008)
  [arXiv:0809.4646 [astro-ph]].
    

\bibitem{Kawasaki:2008pa}
  M.~Kawasaki, K.~Nakayama, T.~Sekiguchi, T.~Suyama and F.~Takahashi,
  JCAP {\bf 0901}, 042 (2009)
  [arXiv:0810.0208 [astro-ph]].


\bibitem{Hikage:2008sk}
  C.~Hikage, K.~Koyama, T.~Matsubara, T.~Takahashi and M.~Yamaguchi,
  Mon.\ Not.\ Roy.\ Astron.\ Soc.\  {\bf 398}, 2188 (2009)
  [arXiv:0812.3500 [astro-ph]].



\bibitem{Kawakami:2009iu}
  E.~Kawakami, M.~Kawasaki, K.~Nakayama and F.~Takahashi,
  JCAP {\bf 0909}, 002 (2009)
  [arXiv:0905.1552 [astro-ph.CO]].
  
   
  
\bibitem{Langlois:2010dz}
  D.~Langlois and A.~Lepidi,
  arXiv:1007.5498v2 [astro-ph.CO], to appear in JCAP. 
  
   
\bibitem{Suyama:2010uj}
  T.~Suyama, T.~Takahashi, M.~Yamaguchi and S.~Yokoyama,
  arXiv:1009.1979 [astro-ph.CO].
 
  

\bibitem{Langlois:2004nn}
  D.~Langlois and F.~Vernizzi,
  Phys.\ Rev.\  D {\bf 70}, 063522 (2004)
  [arXiv:astro-ph/0403258];
  
\bibitem{Moroi:2005kz}
  T.~Moroi, T.~Takahashi and Y.~Toyoda,
  Phys.\ Rev.\  D {\bf 72}, 023502 (2005)
  [arXiv:hep-ph/0501007];
  
\bibitem{Moroi:2005np}
  T.~Moroi and T.~Takahashi,
  Phys.\ Rev.\  D {\bf 72}, 023505 (2005)
  [arXiv:astro-ph/0505339];

\bibitem{Ichikawa:2008iq}
  K.~Ichikawa, T.~Suyama, T.~Takahashi and M.~Yamaguchi,
  Phys.\ Rev.\  D {\bf 78}, 023513 (2008)
  [arXiv:0802.4138 [astro-ph]].
  
\bibitem{Lemoine:2009is}
  M.~Lemoine, J.~Martin and J.~Yokoyama,
  Phys.\ Rev.\  D {\bf 80}, 123514 (2009)
  [arXiv:0904.0126 [astro-ph.CO]].



   
   \bibitem{Langlois:2005ii}
  D.~Langlois and F.~Vernizzi,
  Phys.\ Rev.\ Lett.\  {\bf 95}, 091303 (2005)
  [arXiv:astro-ph/0503416].

\bibitem{Langlois:2005qp}
  D.~Langlois and F.~Vernizzi,
  Phys.\ Rev.\ D {\bf 72}, 103501 (2005)
  [arXiv:astro-ph/0509078].

\bibitem{Langlois:2010vx}
  D.~Langlois and F.~Vernizzi,
  Class.\ Quant.\ Grav.\  {\bf 27}, 124007 (2010)
  [arXiv:1003.3270 [astro-ph.CO]].


\bibitem{Lyth:2004gb}
  D.~H.~Lyth, K.~A.~Malik and M.~Sasaki,
  JCAP {\bf 0505}, 004 (2005)
  [arXiv:astro-ph/0411220].


\bibitem{Enqvist:2005pg}
  K.~Enqvist and S.~Nurmi,
  JCAP {\bf 0510}, 013 (2005)
  [arXiv:astro-ph/0508573].

\bibitem{Huang:2008zj}
  Q.~G.~Huang,
  JCAP {\bf 0811}, 005 (2008)
  [arXiv:0808.1793 [hep-th]].
  

\bibitem{Enqvist:2009zf}
  K.~Enqvist, S.~Nurmi, G.~Rigopoulos, O.~Taanila and T.~Takahashi,
  JCAP {\bf 0911}, 003 (2009)
  [arXiv:0906.3126 [astro-ph.CO]].


\bibitem{Enqvist:2009ww}
  K.~Enqvist, S.~Nurmi, O.~Taanila and T.~Takahashi,
  JCAP {\bf 1004}, 009 (2010)
  [arXiv:0912.4657 [astro-ph.CO]].

\bibitem{Byrnes:2010xd}
  C.~T.~Byrnes, K.~Enqvist and T.~Takahashi,
  JCAP {\bf 1009}, 026 (2010)
  [arXiv:1007.5148 [astro-ph.CO]].





 
\bibitem{Moroi:2002rd}
  T.~Moroi and T.~Takahashi,
  Phys.\ Rev.\  D {\bf 66}, 063501 (2002)
  [arXiv:hep-ph/0206026].



\bibitem{Lyth:2003ip}
  D.~H.~Lyth and D.~Wands,
  Phys.\ Rev.\  D {\bf 68}, 103516 (2003)
  [arXiv:astro-ph/0306500].


  
   
\bibitem{Beltran:2008ei}
  M.~Beltran,
  Phys.\ Rev.\  D {\bf 78}, 023530 (2008)
  [arXiv:0804.1097 [astro-ph]].
   
\bibitem{Moroi:2008nn}
  T.~Moroi and T.~Takahashi,
  Phys.\ Lett.\  B {\bf 671}, 339 (2009)
  [arXiv:0810.0189 [hep-ph]].

\bibitem{Takahashi:2009cx}
  T.~Takahashi, M.~Yamaguchi and S.~Yokoyama,
  Phys.\ Rev.\  D {\bf 80}, 063524 (2009)
  [arXiv:0907.3052 [astro-ph.CO]].

\bibitem{Sasaki:2006kq}
  M.~Sasaki, J.~Valiviita and D.~Wands,
  Phys.\ Rev.\  D {\bf 74}, 103003 (2006)
  [arXiv:astro-ph/0607627].

\bibitem{Langlois:1999dw}
  D.~Langlois,
  Phys.\ Rev.\ D {\bf 59}, 123512 (1999)
  [arXiv:astro-ph/9906080].


 
\bibitem{Langlois:2000ar}
  D.~Langlois and A.~Riazuelo,
  Phys.\ Rev.\  D {\bf 62}, 043504 (2000)
  [arXiv:astro-ph/9912497].
  
  
  
\bibitem{Komatsu:2008hk}
  E.~Komatsu {\it et al.}  [WMAP Collaboration],
  Astrophys.\ J.\ Suppl.\  {\bf 180}, 330 (2009)
  [arXiv:0803.0547 [astro-ph]].


\bibitem{Lyth:2005fi}
  D.~H.~Lyth and Y.~Rodriguez,
  Phys.\ Rev.\ Lett.\  {\bf 95}, 121302 (2005)
  [arXiv:astro-ph/0504045].


\bibitem{Byrnes:2006vq}
  C.~T.~Byrnes, M.~Sasaki and D.~Wands,
  Phys.\ Rev.\  D {\bf 74}, 123519 (2006)
  [arXiv:astro-ph/0611075].


\bibitem{Smidt:2010sv}
  J.~Smidt, A.~Amblard, A.~Cooray, A.~Heavens, D.~Munshi and P.~Serra,
  arXiv:1001.5026 [astro-ph.CO].
  
  

 \end{thebibliography}
\end{document}